\documentclass[aps,prb,twocolumn,superscriptaddress,showpacs,floatfix]{revtex4}

\usepackage{graphicx}

\begin{document}

\title{Asymmetry of magnetic-field profiles in
       superconducting strips}

\author{G.~P.~Mikitik}
\affiliation{Max-Planck-Institut f\"ur Metallforschung,
   D-70506 Stuttgart, Germany}
\affiliation{B.~Verkin Institute for Low Temperature Physics
   \& Engineering, Ukrainian Academy of Sciences,
   Kharkov 61103, Ukraine}

\author{E.~H.~Brandt}
\affiliation{Max-Planck-Institut f\"ur Metallforschung,
   D-70506 Stuttgart, Germany}

\date{\today}

\begin{abstract}
We analyze the magnetic-field profiles $H_z(x)$ at the upper
(lower) surface of a superconducting strip in an external magnetic
field, with $z$ perpendicular to the plane of the strip and $x$
along its width. The external magnetic field $H_a$ is
perpendicular or inclined to the plane of the strip. We show that
an asymmetry of the profiles $H_z(x)$ appears in an oblique
magnetic field $H_a$ and also in the case when the angular
dependence of the critical current density $j_c$ in the
superconductor is not symmetric relative to the $z$ axis. The
asymmetry of the profiles is related to the difference $\Delta
H_z(x)$ of the magnetic fields at the upper and lower surfaces of
the strip, which we calculate. Measurement of this difference or,
equivalently, of the asymmetry of the profiles can be used as a
new tool for investigation of flux-line pinning in
superconductors.
\end{abstract}


\maketitle

\section{Introduction}

In a recent Letter \cite{t} magnetic-field profiles at the upper
surface of a thin rectangular YBa$_2$Cu$_3$O$_{7-\delta}$ platelet
placed in a perpendicular external magnetic field were
investigated by magneto-optical imaging, and the following
interesting observation was made: When columnar defects slightly
tilted to the c-axis (normal to the platelet surface) were
introduced into the sample by heavy-ion irradiation, an asymmetry
of the magnetic-field profiles relative to the central axis of the
sample appeared, and this asymmetry nonmonotonically depended on
the magnitude of the external magnetic field $H_a$, disappearing
at large $H_a$. The authors of Ref.~\onlinecite{t} explained the
asymmetry by in-plane magnetization originating from a zigzag
structure of vortices, i.e., from their partial alignment along
the columnar defects. They also implied that when the vortices
lose their interlayer coherence, the in-plane magnetization and
hence the asymmetry disappear. On this basis, it was claimed
\cite{t} that the asymmetry can be a powerful probe for the
interlayer coherence in superconductors. In this paper we show
that the asymmetry of the magnetic field profiles in thin flat
superconductors may have a more general origin. It may result from
anisotropy of flux-line pinning and needs not be due to the kinked
structure of vortices. The asymmetry can occur not only in layered
superconductors but also in three dimensional materials, and its
disappearance can be understood without the assumption that the
interlayer coherence is lost. Interestingly, without any columnar
defects, such an asymmetry of the magnetic field profiles was
recently observed in a Nb$_3$Sn slab placed in an oblique magnetic
field. \cite{w}

In Ref.~\onlinecite{1,obl1} we explained how to solve the critical
state problem for thin flat three-dimensional superconductors with
an arbitrary anisotropy of flux-line pinning. But these equations
yield magnetic-field profiles in the critical state only to the
leading order in the small parameter $d/w$ where $d$ is the
thickness of the flat superconductor and $w$ is its characteristic
lateral dimension. In this approximation the magnetic-field
component perpendicular to the flat surfaces of the sample, $H_z$,
is independent of the coordinate $z$ across the thickness of the
superconductor and coincides with the appropriate magnetic field
of an infinitely thin superconductor of the same shape (but with
some dependence of the critical sheet current $J_c$ on $H_z$). As
it will be seen below, the asymmetry of the profiles is related to
the difference $\Delta H_z$ of the fields $H_z$ at the upper and
lower surfaces of the sample. Thus, to describe the asymmetry of
the $H_z$ profiles, it is necessary to consider these profiles
more precisely, to the next order in $d/w$, taking into account
the dependence of $H_z$ on the coordinate across the thickness of
the superconductor.

In this paper we obtain formulas for $\Delta H_z$ and for the
asymmetry of the magnetic-field profiles in an infinitely long
thin strip, and discuss the conditions under which the asymmetry
can be observed. In particular, an asymmetry always will appear
for strips in an oblique magnetic field, and experimental
investigation of this asymmetry provides new possibilities for
analyzing flux-line pinning in superconductors. We also
demonstrate that the experimental data of Ref.~\onlinecite{t} can
be qualitatively understood by assuming some anisotropy of pinning
in a three-dimensional (not layered) superconductor.

\section{Magnetic-field profiles of strips}

In this paper we consider the following situation: A thin three
dimensional superconducting strip fills the space $|x| \le w$,
$|y| < \infty$, $|z| \le d/2$ with $d\ll w$; a constant and
homogeneous external magnetic field $H_a$ is applied at an angle
$\theta_0$ to the $z$ axis ($H_{ax}=H_a\sin\theta_0$, $H_{ay}=0$,
$H_{az}=H_a\cos\theta_0$). For definiteness, we shall imply below
that $H_{ax}$ is switched on first and then $H_{az}$ is applied,
i.e., the so-called third scenario \cite{obl1} of switching on
$H_a$ occurs (for the definition of the first and second scenarios
see Appendix B). It is also assumed that surface pinning is
negligible, the thickness of the strip, $d$, exceeds the London
penetration depth, and the lower critical field $H_{c1}$ is
sufficiently small so that we may put $B= \mu_0 H$.

The symmetry of the problem leads to the following relationships:
\[
j_y(x,z)=-j_y(-x,-z),
\]
\[
H_z(x,z)=H_z(-x,-z),\ \ \  H_x(x,z)=H_x(-x,-z),
\]
where $j_y(x,z)$ is the current density flowing at the point
($x$,$z$). In other words, the field at the lower surface of the
strip, $H_z^-(x)\equiv H_z(x,-d/2)$, can be expressed via the
field at its upper surface, $H_z^+(x)\equiv H_z(x,d/2)$, as
follows:
 \[
 H_z^-(x)=H_z^+(-x),
 \]
and hence for the difference $\Delta H_z(x)\equiv H_z^+(x) -
H_z^-(x)$ of the fields at the upper and lower surfaces we obtain
the formula
\begin{equation} \label{1}
\Delta H_z(x)=H_z^+(x)-H_z^+(-x) ,
\end{equation}
which connects this $\Delta H_z(x)$ and the asymmetry of the
magnetic-field profile at the upper surface of the strip.

As mentioned in the Introduction, to leading order in $d/w$ the
critical state problem for such a strip can be reduced to the
critical state problem for the infinitely thin strip with some
dependence of the critical sheet current $J_c$ on $H_z$ where the
sheet current is the current density integrated over the thickness
of the strip,
\[
 J \equiv \int_{-d/2}^{d/2} j_y(x,z)dz ,
\]
and $J_c$ is its critical value. The dependence $J_c(H_z)$ results
from both a dependence of the critical current density $j_c$ on
the absolute value of the local magnetic induction $\mu_0|H|$ and
an out-of-plane anisotropy of $j_c$, i.e., a dependence of $j_c$
on the angle $\theta$ between the local direction of ${\bf H}$ and
the $z$ axis. Since both $|H|$ and $\theta$ change with $z$ in
strips of finite thickness, this means that $J_c\neq j_cd$, and a
dependence of $J_c$ on $H_z$ appears. The function $J_c(H_z)$ can
be found from the equation, \cite{1,obl1}
\begin{equation} \label{1a}
 d= \!\! \int_{H_x^-}^{H_x^+}\!\!{dh
 \over j_c(h,H_z)} ,
\end{equation}
where the critical current density $j_c(H_x,H_z)$ may have
arbitrary dependence on the local $H_x$ and $H_z$;
$H_x^-=H_{ax}-0.5J_c(H_z)$, and $H_x^+=H_{ax}+0.5J_c(H_z)$ are the
$x$ component of the magnetic field at the lower and the upper
surfaces of the strip. The function $J_c(H_z)$ found from
Eq.~(\ref{1a}) generally depends on the parameter $H_{ax}$, and
only within the Bean model when $j_c$ is independent of ${\bf H}$,
equation (\ref{1a}) yields $J_c=j_cd$ for any $H_{ax}$.

Formula (\ref{1a}) is valid to the leading order in $d/w$ since we
neglected the term $\partial H_z(x,z)/ \partial x$ in the equation
${\rm rot}{\bf H}={\bf j}$ and used the expression
\begin{equation} \label{4}
 {\partial H_x(x,z)\over \partial z}= j_y(x,z)
\end{equation}
with $j_y(x,z)=j_c(H_x,H_z)$ to derive formula (\ref{1a}). In this
approximation $H_z$ is independent of $z$ inside the strip. Thus
$\Delta H_z(x)=0$, and so the $H_z$ profile is always symmetric.
This profile is given by the Biot-Savart law for the infinitely
thin strip
\begin{equation} \label{2}
 H_z(x)=H_{az}+{1 \over 2\pi}\!\int_{-w}^w\!\!
 { J(t)\, dt \over t-x}  \,.
\end{equation}
Here the sheet current $J(x)$ follows from the critical state
equations for this strip:
\begin{equation}\label{C3}
 J(x)=-{x\over |x|}J_c[H_z(x)]
\end{equation}
if $a \le |x| \le w$, and
\begin{equation}\label{C4}
 H_z(x)=0
\end{equation}
when $|x|\le a$. The points $x=\pm a$ give the position of the
flux front in the infinitely thin strip. The integral equations
(\ref{2}) - (\ref{C4}) can be solved by either a static iterative
method, or more conveniently by a dynamic method. \cite{EH,EH1}

An analysis of $H_z$ allowing for terms of the order of $d/w$ is
presented in Appendix A. In this case $\Delta H_z\neq 0$, and it
is given by
\begin{eqnarray} \label{5}  
 \Delta H_z(x) = {d\over dx}
 \int_{-d/2}^{d/2} zj_y(x,z) dz  \,.
\end{eqnarray}
This expression can be also derived from the following simple
considerations: Using ${\rm div}{\bf H}=0$, we write
\begin{equation} \label{3}  
 \Delta H_z(x)=\!\! \int_{-d/2}^{d/2}\!\!{\partial H_z(x,z)\over \partial
 z} dz = -\! \!\int_{-d/2}^{d/2}\!\!{\partial H_x(x,z)\over \partial x}
 dz .
\end{equation}
From Eq.~(\ref{4}) it follows that
\[
 H_x(x,z)=H_x(x,-d/2)+\int_{-d/2}^z j_y(x,z') dz' .
\]
Inserting this expression into formula (\ref{3}), interchanging
the sequence of the integrations, and using
$H_x(x,-d/2)=H_{ax}-0.5J(x)$, we find formula (\ref{5}). It
follows from this formula that an asymmetry of the $H_z$-profile
can appear only if the distribution of the current density $j_y$
across the thickness of the strip is asymmetric about the middle
plane of the strip, $z=0$, and if this distribution changes with
$x$ (the latter condition was not obtained in
Ref.~\onlinecite{t}). \cite{c1}

\section{Conditions for asymmetry}

If the angular dependence of the critical current density is
symmetric relative to the $z$ axis, $j_c(-H_x,H_z)=j_c(H_x,H_z)$,
an asymmetry of the current distribution across the thickness of
the strip can occur only in an oblique applied magnetic field
which breaks the relation $H_x(x,-z)=-H_x(x,z)$. Besides this,
asymmetry of the distribution can appear for asymmetric angular
dependence of $j_c$, $j_c(-H_x,H_z)\neq j_c(H_x,H_z)$, even when
the external magnetic field is applied along the $z$ axis. In this
section we consider the case of an oblique magnetic field.

\subsection{Region where $H_z\approx 0$}

\begin{figure}  
\includegraphics[scale=.45]{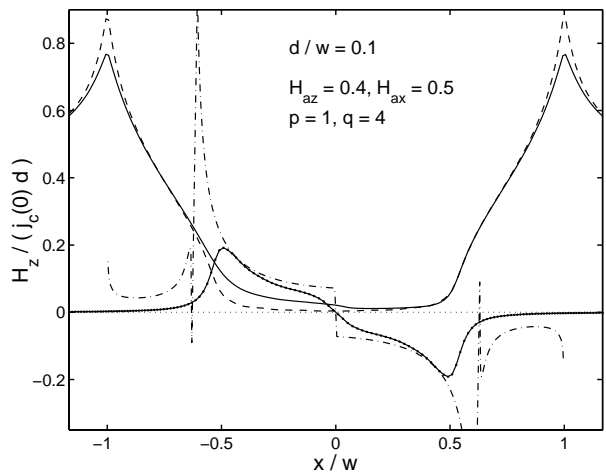}
\caption{\label{fig1} Asymmetry of the magnetic field profile
$\Delta H_z(x)$ at the upper surface of a thin strip with $d=0.1w$
and $j_c(\theta)$ given by Eq.~(\ref{12}) with $p=1$, $q=4$, see
Fig.~2. Here $H_{az}=0.4$, $H_{ax}=0.5$; the magnetic fields are
measured in units of $j_c(0)d$. The dash-dot line shows $3\Delta
H_z(x)$ calculated with Eq.~(\ref{11a}) and formulas of Appendix
B. The discontinuities of this line seen at $x=0$, $x=\pm a$, and
$x=\pm w$ are due to inapplicability of our approximation in the
vicinity of these points. For comparison, the solid line with dots
shows $3\Delta H_z(x)$ calculated directly by solving the
two-dimensional critical state problem \cite{EH1} (creep exponent
$\sigma = 200 \gg 1$; London depth $\lambda = 0.1 d$). The dashed
line and the solid line without dots give the profiles $H_z(x,0)$
and $H_z(x,d/2)$, obtained directly from this solution of the
two-dimensional problem.
 } \end{figure}

For a superconducting strip in an oblique magnetic field, it was
shown recently \cite{mbi} that in the region of the strip, $|x|\le
a$, where a flux-free core occurs, i.e., where $H_z\approx 0$, the
distribution of the current across the thickness of the sample is
highly asymmetric even for superconductors with ${\bf
H}$-independent $j_c$ (the Bean model). Using this distribution
(which depends on how the magnetic field is switched on) and
Eq.~(\ref{5}), one can calculate $\Delta H_z(x)$ in this region of
the strip. Note that the results of such calculations may be also
applied to anisotropic strips with $j_c=j_c(\theta)$ since in this
region of the strip the flux lines practically lie in the $x$-$y$
plane, $j_c$ is independent of the coordinates, $j_c\approx
j_c(\pi/2)$, and the results for the flux-free core obtained
within the Bean model remain applicable to this anisotropic case.
In Fig.~\ref{fig1} we present this $\Delta H_z(x)$ for the
anisotropic strip in the case of the third scenario \cite{obl1} of
switching on $H_a$ when the field $H_{ax}$ is switched on first
and then the component $H_{az}$ is applied, see Appendix B. Thus,
a nonzero $\Delta H_z$ in this region of the sample reflects the
asymmetry of the flux-free core, and this $\Delta H_z$ differs
from zero in an oblique magnetic field for any superconductor.

The quantity $\Delta H_z(x)$ in the region $|x| < a$ depends on
how the external magnetic field is switched on. In particular,
when $H_{ax}$ is applied before $H_{az}$ (the third scenario) and
$j_c d/2 < H_{ax}$, $\Delta H_z(x)$ is described by Eq.~(B9). On
the other hand, for the same applied field but with $H_{ax}$ and
$H_{az}$ switched on simultaneously (the so-called first scenario
\cite{mbi}), we obtain from formulas of Ref.~\onlinecite{mbi} at
$|x|<a \sin\theta_0$
\begin{equation}\label{B9a}   
 \Delta H_z(x)=-{J(x)\over 2j_c}{dJ(x)\over dx}.
\end{equation}
Comparison of Eqs.~(B9) and (\ref{B9a}) demonstrates that
measurements of the asymmetry of the $H_z$ profiles at the upper
surface of the strip enable one to investigate subtle differences
between critical states generated by different scenarios of
switching on $H_a$ (to the leading order in the small parameter
$d/w$ one has $H_z=0$ at $|x|<a$ for any scenario). Note also that
formulas of type (B9) [or (\ref{B9a})] permit one to find the
sheet current $J(x)$ in the region $|x|<a$ from $\Delta H_z$.

\subsection{Region where $H_z\neq 0$}

Consider now the region of the strip penetrated by $H_z$, $a\le
|x| \le w$, i.e., the region where the component $H_z$ differs
from zero. In this case it is useful to represent Eq.~(\ref{5}) in
another form, using the solution $H_x(z, H_z(x))$ of
Eq.~(\ref{4}). This solution is implicitly given by
\begin{equation} \label{6}
 z+(d/2)= \!\!\pm \int_{H_x^{\mp}}^{H_x} {dh\over j_c(h,H_z)} ,
\end{equation}
where $H_x^-$ and $H_x^+$ are the same as in Eq.~(\ref{1a}). Using
Eq.~(\ref{6}), we can transform the integration over $z$ in
Eq.~(\ref{5}) into an integration over $H_x$. After a simple
manipulation, we obtain
\begin{eqnarray} \label{7}
 \Delta H_z(x) = - {d\over dx}
 \left (\int_{H_x^-}^{H_x^+} {h\,dh
 \over j_c(h,H_z)} \right ) \,.
\end{eqnarray}
It is clear from this formula that $\Delta H_z$ depends on $x$ via
the function $H_z(x)$, i.e., $\Delta H_z$ has the form: $\Delta
H_z=d\cdot (dH_z/dx)F(H_z)$ where $F$ is some dimensionless
function of $H_z$.

Within the Bean model when $j_c$ is independent of ${\bf H}$ and
$J_c=j_cd$, equation (\ref{7}) yields $\Delta H_z(x)=0$, and hence
the asymmetry of the profiles $H_z(x)$ in the fully penetrated
region of the strip is absent for any $H_{ax}$. The asymmetry
appears only if there is a dependence of $j_c$ on $|{\bf H}|$ or
if there is an anisotropy of pinning (or it may result from both
reasons). If $j_c(H_x,H_z)=j_c(-H_x,H_z)$, i.e., if the angular
dependence of the critical current density is symmetric relative
to the $z$ axis, it follows from formula (\ref{7}) that $\Delta
H_z(x)=0$ at $H_{ax}=0$. In other words, for such $j_c(H_x,H_z)$
the asymmetry of the magnetic-field profiles can appear only in an
oblique magnetic field. At small $H_{ax}$, Eq.~(\ref{7}) yields
\begin{eqnarray} \label{8}  
 {\Delta H_z(x)\over H_{ax}} = - {d\over dx}
 \left ( {J_c(H_z) \over j_c(H_x,H_z)} \right ) \,,
\end{eqnarray}
where $H_x=J_c(H_z)/2$, and $J_c(H_z)$ is the dependence of the
critical sheet current on $H_z$ at $H_{ax}=0$. The integration of
this formula leads to the relationship
\begin{eqnarray} \label{9}  
 \int^{x_0}\!\!\!dx \left ({\Delta H_z(x)\over H_{ax}}\right ) =
 {\rm const}- {J_c(H_z) \over j_c(H_x,H_z)}  \,,
\end{eqnarray}
which, in principle, enables one to reconstruct the function
$j_c(H_x,H_z)$ (up to a constant) if the function $\Delta H_z(x)$
at small $H_{ax}$ and the functions $H_z(x)$ and $J_c(H_z)$ at
$H_{ax}=0$ are known. Here $H_x=J_c(H_z)/2$, and $H_z=H_z(x_0)$.
The integration in Eq.~(\ref{9}) is carried out over the region
where $H_z\neq 0$.

It was shown recently \cite{obl1} that in the case of ${\bf
H}$-dependent $j_c$ the magnetic field profiles at the upper
surface of the strip generally depend on the scenario of switching
on the oblique magnetic field. In Ref.~\onlinecite{obl1} the
profiles were analyzed to the leading order in $d/w$, and so they
were always symmetric in $x$, $H_z(-x,d/2)=H_z(x,d/2)$. Formula
(\ref{7}), which describes the antisymmetric part of the profiles
(this part appears to the next order in $d/w$), has been derived
under the assumption that the sign of $j_y$ remains unchanged
across the thickness of the strip. This assumption is indeed valid
for the third scenario discussed in this paper. However, there
exist scenarios, see, e.g., Ref.~\onlinecite{obl1}, when
$j_y(x,z)$ changes its sign at some boundary $z=z_c(x)$. It is
this boundary that causes the difference between the profiles for
the different scenarios. Of course, the existence of this boundary
also implies a modification of formula (\ref{7}). Thus, we expect
that if for some scenarios the symmetric parts of the $H_z$
profiles differ, their antisymmetric parts have to differ, too.

  The magnetic-field profiles at the upper surface of the
strip are obtained either by magneto-optics, see, e.g.,
Ref.~\onlinecite{In,Kob,Kr,4}, or using Hall-sensor
arrays.\cite{Z} These profiles enable one to find the
sheet-current distribution $J(x)$,\cite{R,JH,Gr,W,J,G,Joo} and
then the dependence $J_c(H_z)$. \cite{4} In Ref.~\onlinecite{obl1}
we discussed a way how to determine $j_c$ from $J_c(H_z)$. In this
context, measurements of $\Delta H_z(x)$ in oblique magnetic
fields can provide additional information on the flux-line pinning
in superconductors when the critical current density $j_c$ depends
on the magnitude or direction of the magnetic field.

  To demonstrate this, we compare $\Delta H_z(x)$ for two types of
pinning: For the first type the critical current density depends
only on the combination $|H|\cos\theta=H_z$ where $\theta$ is the
angle between the local direction of ${\bf H}$ and the $z$ axis;
for the second type $j_c$ is a function of $\theta$ only. The
first situation occurs for the case of weak collective pinning by
point defects in the small bundle pinning regime when the scaling
approach is valid. \cite{bl} In this case $j_c$ depends on the
combination $|H|(\cos^2\theta +\epsilon^2\sin^2\theta)^{1/2}$
which practically coincides with $|H|\cos\theta=H_z$ if the
anisotropy parameter $\epsilon$ is small. Then, equation (\ref{7})
gives
 \[
  \Delta H_z(x)=-H_{ax}{d\over dx}\left ({J_c(H_z)\over j_c(H_z)}
  \right ) =0,
 \]
i.e., the asymmetry is absent in the region $|x|>a$ for this type
of pinning. Here we have taken into account that in this situation
Eq.~(\ref{1a}) reduces to $J_c(H_z)=dj_c(H_z)$. We are coming now
to an analysis of $\Delta H_z(x)$ for the second type of pinning.

\subsection{$j_c$ depends only on the direction of
{\bf H}}

Let us consider more closely the case when $j_c$ depends only on
$\theta$, $j_c=j_c(\theta)$, where $\theta$ is the angle between
the local direction of ${\bf H}$ and the $z$ axis. In other words,
we shall analyze the situation when the dependence of $j_c$ on
$|H|$ is negligible. This approximation can be justified for not
too thick samples with anisotropic pinning. \cite{obl1} Here we
also assume the symmetry $j_c(\theta)=j_c(-\theta)$. In this case,
one can express the dependence of the critical current density
$j_c$ on $\theta$ and the dependence of $J_c$ on $H_z$ at
$H_{ax}\neq 0$ in terms of the function $J_c(H_z)$ at $H_{ax}=0$,
\cite{1,obl1} see Appendix C. The quantity $\Delta H_z(x)$,
Eq.~(\ref{7}), is also expressible in terms of $J_c(H_z,0)$, the
sheet current $J_c(H_z)$ at $H_{ax}=0$,
\begin{eqnarray} \label{11a}
 \Delta H_z(x) = -d\,{dH_z\over dx}{d \over dH_z}
\! \left [\! H_z^2 \int_{t_+}^{t_-}{J_c(t,0)dt \over 4t^3}\!
\right ]\!.
\end{eqnarray}
Here we have used the parametric representation (\ref{2a}), and
formulas (\ref{10}) that determine the auxiliary variables $t_+$
and $t_-$. As to Eq.~(\ref{8}), it takes the form:
 \begin{eqnarray} \label{11b}
{\Delta H_z(x)\over H_{ax}} = -d\,{dH_z\over dx}{d \over dH_z} \!
\left [\! {J_c(H_z) \over J_c(H_z)-H_z(dJ_c/dH_z)}\! \right]\!,
 \end{eqnarray}
where $J_c(H_z)$ is the $H_z$ dependence of $J_c$ at $H_{ax}=0$.

 \begin{figure}  
\includegraphics[scale=.45]{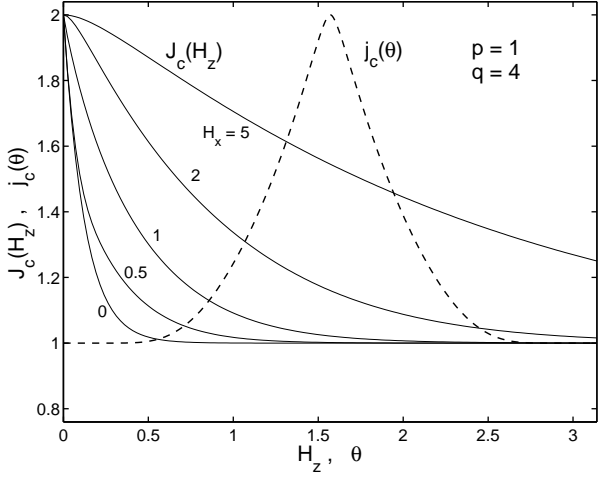}
\caption{\label{fig2} Angular dependence of the critical current
density $j_c(\theta)$, Eq.~(\ref{12}), for $p=1$, $q=4$ (dashed
line). The solid lines show the corresponding dependences of the
sheet current $J_c$ on $H_z$ at $H_{ax}=0$, Eq.~(\ref{11}), and at
$H_{ax}=0.5$, $1$, $2$, $5$. The dependences  for nonzero $H_{ax}$
are obtained from Eqs.~(\ref{10}), (\ref{11}). The current density
is measured in units of $j_c(0)$, while $J_c$ and $H_z$ are in
units of $j_c(0)d$.
 } \end{figure}  

 \begin{figure}  
\includegraphics[scale=.50]{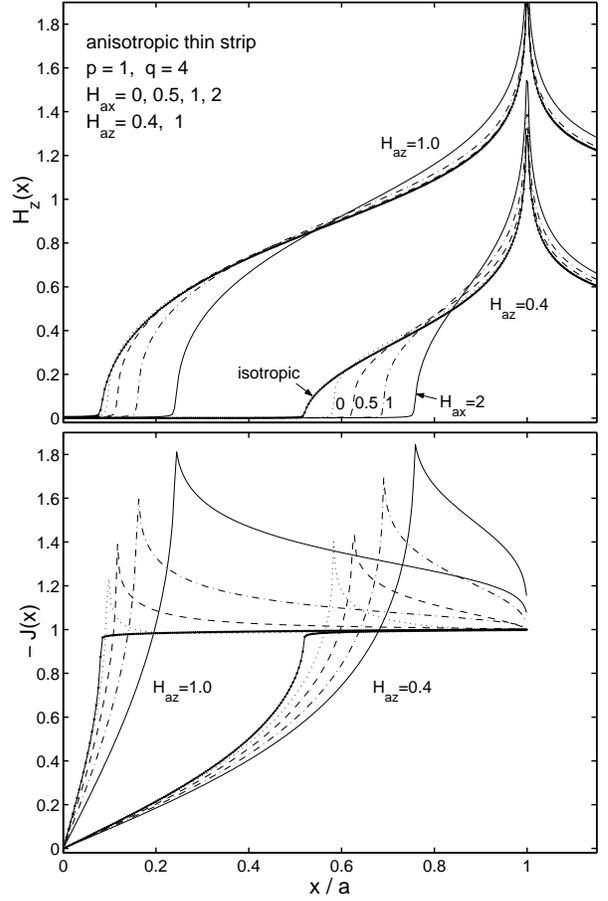}
\caption{\label{fig3} Spatial profiles of the perpendicular field
component  $H_z(x)$ (upper plot) and of the sheet current $J(x)$
(lower plot) of a thin strip with anisotropic pinning described by
model (\ref{12}) with $p=1$, $q=4$, see  Fig.~2. The various
curves correspond to increasing applied field $H_{az}=0.4$ and $1$
in units of $j_c(0)d$.  The dotted, dashed,  dot-dashed, and solid
curves are for $H_{ax}=0$, 0.5, 1, and 2, respectively. For
comparison, the solid curves with dots (indicating the grid) show
the profiles for isotropic pinning ($p=0$).
 } \end{figure}  

 \begin{figure}  
\includegraphics[scale=.49]{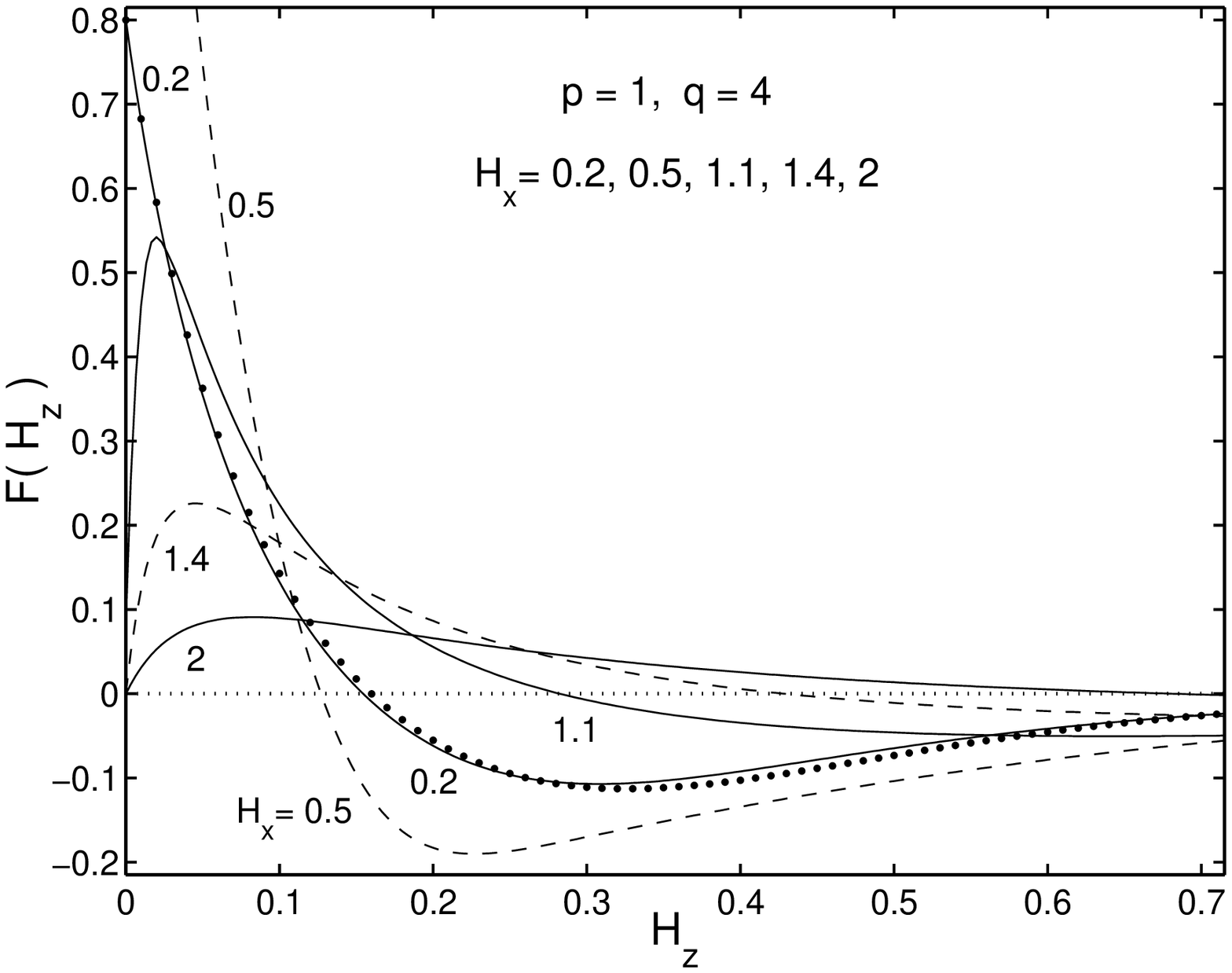}
\caption{\label{fig4} The function $F(H_z)\equiv \Delta
H_z(x)(dH_z/dx)^{-1}d^{-1}$ calculated with Eqs.~(\ref{11a}),
(\ref{11}) for $p=1$, $q=4$ and various $H_{ax}=0.2$, $0.5$,
$1.1$, $1.4$, $2$. The dots show the $F(H_z)$ calculated from
Eq.~(\ref{11b}) for $H_{ax}=0.2$. $H_z$ and $H_{ax}$ are measured
in units of $j_c(0)d$.
 } \end{figure}  

We now present an example of such calculations for this type of
pinning. Let at $H_{ax}=0$ the following dependence $J_c(H_z)$ be
extracted from some experimental magneto-optics data:
    \begin{eqnarray}  \label{11}  
    J_c(H_z) = j_c(0) d \left [ 1 + p\, \exp\left
    (-{q \,H_z \over H_{cr}}\right ) \right ] \,,
    \end{eqnarray}
where $H_{cr} = j_c(0) d /2$, while $j_c(0)$ and the dimensionless
$p$ and $q$ are positive constants. Using Eqs.~(\ref{2a}), one can
easily verify that the corresponding angular dependence of the
critical current density takes the form:
     \begin{eqnarray}  \label{12}  
     j_c(\theta) =j_c(0)\left [ 1 + p \, (1+ q\, t)
     \exp(- q\, t)\right ] \,, \nonumber \\
     \tan \theta = t^{-1}\left [ 1 + p\,
     \exp(- q\, t)\right ] \,,
     \end{eqnarray}
where $t$ is a curve parameter with range $0 \le t \le \infty$.
This dependence $j_c(\theta)$ is presented in Fig.~\ref{fig2}
together with the function $J(H_z)$, Eq.~(\ref{11}). Note that for
$p>0$ the character of this dependence $j_c(\theta)$ is typical of
layered high-$T_c$ superconductors, \cite{1} i.e., $j_c$ is
largest for $\theta=\pi/2$. The profiles $H_z(x)$ that correspond
to this $J_c(H_z)$ are shown in Fig.~\ref{fig3}. In Figs.~2 and 3
we also show the dependences of $J_c$ on $H_z$ for $H_{ax}\neq 0$
and appropriate magnetic-field profiles in an oblique magnetic
field. With the use of Eqs.~(\ref{10}), (\ref{11a}) - (\ref{11}),
we calculate the quantity $(\Delta H_z/d)(dH_z/dx)^{-1}$ as a
function of $H_z$, Fig.~\ref{fig4}. Interestingly, even at not too
small $H_{ax}< 0.4j_c(0)d$, formula (\ref{11b}) gives reasonable
results. An example of the asymmetry of the $H_z$-profiles at the
upper surface of the strip is presented in Fig.~1. This asymmetry
is found from the data of Fig.~4, the known derivative $dH_z(x)/dx$,
Fig.~3, and formulas of Appendix B (when $|x|<a$). Note that the
steepness of $H_z(x)$ near $x=a$ has a pronounced effect on the
form of $\Delta H_z(x)$ (and this steepness essentially depends
\cite{1} on the sign of $p$). The discontinuities of $\Delta
H_z(x)$ at $x=\pm a$, $0$, $\pm w$ are caused by the
inapplicability of our formulas for $\Delta H_z$ there. For
comparison, we also show $\Delta H_z(x)$ and the profiles
$H_z(x,0)$ and $H_z(x,d/2)$ calculated directly by solving the
two-dimensional critical state problem for a strip of finite
thickness.\cite{EH1} In this two-dimensional calculation the
discontinuities of $\Delta H_z$ are smoothed out on scales of the
order of $d$.

\section{Strip with inclined defects in perpendicular
magnetic field}  

In a recent Letter \cite{t} the position of the so-called central
d-line (the discontinuity line) at the upper surface of a thin
rectangular YBa$_2$Cu$_3$O$_{7-\delta}$ platelet was measured by
magneto-optical imaging. In this d-line the sheet current changes
its sign and the magnetic field $H_z$ reaches an extremum (a
minimum). When columnar defects tilted to the c-axis were
introduced into the sample, the d-line shifted relative to the
central axis of the platelet, \cite{t} and the value of this shift
first increased and then decreased with increasing perpendicular
magnetic field. As mentioned in the Introduction, the
authors of Ref.~\onlinecite{t} explained the shift by the in-plane
magnetization originating from a zigzag structure of vortices,
and the disappearance of the magnetization and of this shift by
loss of their interlayer coherence. It is clear that the shift of
the d-line reflects the asymmetry of the magnetic field profiles
at the upper surface of the sample.

 \begin{figure}  
\includegraphics[scale=.5]{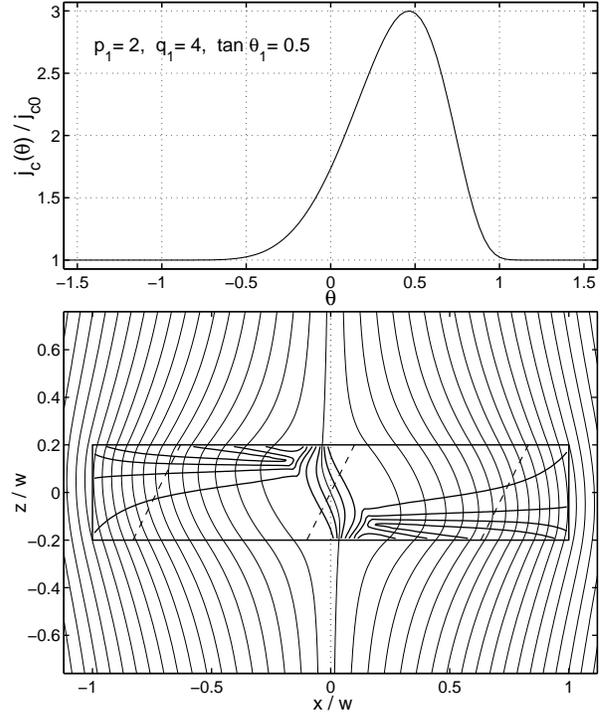}
\caption{\label{fig5} Top: The anisotropic critical current
density $j_c(\theta)$, Eq.~(\ref{13}) with  $p_1 = 2$, $q_1 = 4$,
and $\tan\theta_1 =0.5$, that models pinning by tilted columnar
defects. Bottom: Two-dimensional computation \cite{EH1} (creep
exponent $\sigma = 50 \gg 1$; London depth $\lambda = w/30 =
d/12$; $120\times 28$ grid points) of the critical state in a long
strip with rectangular cross section of aspect ratio $d/2w = 0.2$
and with anisotropic pinning described by the $j_c(\theta)$ of the
upper plot, exposed to a perpendicular field $H_{az} = 0.2
j_{c0}d$. Shown are the magnetic field lines (thin lines) and some
contour lines of the current density $j(x,z) / j_{c0}  =$ -2.75,
-2.25, -1.75, $\dots$, 2.75 (thick lines). The three inclined
dashed lines indicate the direction $\theta_1$ of the columnar
pins. Note that the distribution of $j_c$ across the thickness of
the strip is asymmetric; $|j_c(x,z)|$ is maximum where the field
lines are along the defects, and $j_c \approx j_{c0}$ is nearly
constant where the deviation from this direction is large. The
line $j_c(x,z)=0$ coincides with the central magnetic field line.
 } \end{figure}  

The zigzag structure of vortices occurs when the tilt angle
$\Delta \theta$ of the local ${\bf H}$ to the direction of the
columnar defects is less than the so-called trapping angle
$\theta_t$. \cite{bl1} In this case a misalignment of the local
${\bf H}$ and the averaged direction of the kinked flux lines
appears, and this misalignment is of the order of
$\epsilon^2(H_{c1}/H)(\theta_t- \Delta \theta)$ where $H_{c1}$ is
the lower critical field and $\epsilon$ is the anisotropy
parameter of the superconductor. The misalignment generates an
in-plane magnetization $M_x \sim \epsilon^2 H_{c1} (\theta_t-
\Delta \theta)$, and hence one may expect that $\Delta H_z \approx
-(d/dx) \int_{-d/2}^{d/2}M_xdz$; see Eq.~(\ref{5}). Note that if
$\epsilon \to 0$ (the interlayer coherence is lost), $\Delta H_z$
tends to zero. This is just the mechanism of the asymmetry
discussed in Ref.~\onlinecite{t}, and in this consideration the
asymmetry is due to the equilibrium part of the in-plane
magnetization. However, in our approximation, when $H \gg H_{c1}$
and ${\bf B}=\mu_0{\bf H}$, we neglect the misalignment, the
fields of the order of $H_{c1}$, and the equilibrium part of
magnetization. In our approach we take into account only the
nonequilibrium part of magnetization, which also generates an
asymmetry. This asymmetry can occur even if $H_{c1} \to 0$.

\begin{figure}  
\includegraphics[scale=.485]{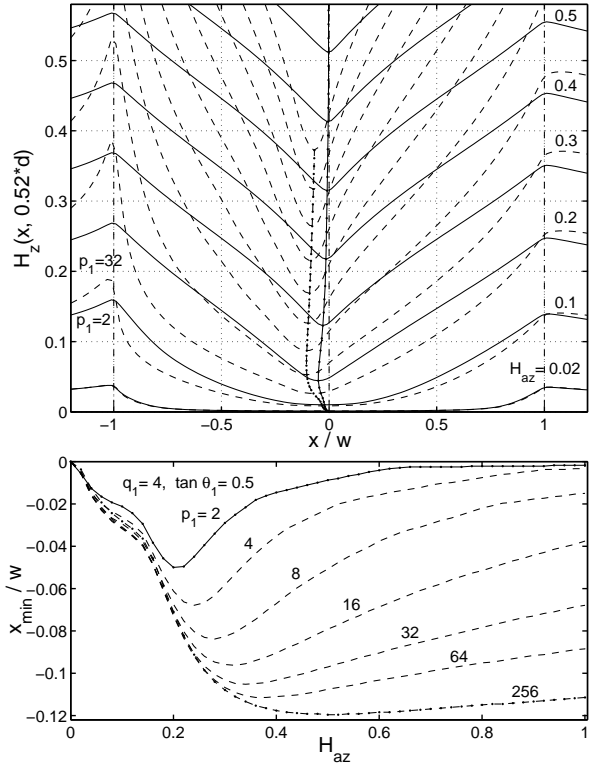}  
\caption{\label{fig6} Top: Asymmetric profiles of the magnetic
field component $H_z(x, z)$ in the plane $z = 0.52 d$ slightly
above the surface of the strip of Fig.~5 with anisotropic pinning
by tilted columnar defects. Shown are the cases $p_1=2$ (solid
lines) and $p_1=32$ (dashed lines), with $q_1=4$, $\tan\theta
=0.5$, at applied field values $H_{az}=0.02$, $0.1$, $0.2$, $0.3$,
$\dots$. Field unit is $j_{c0}d$. The slightly curved, nearly
vertical lines with dots are the locus of the minima of these
$H_z$ profiles. Bottom: The position $x_{\rm min}$ of the minimum
of the profiles $H_z(x, 0.52d)$ plotted versus the applied field
$H_{az}$ for $p_1=2$ $\dots$ 256.
 } \end{figure}

In Sec.~III we have considered the case when the angular
dependence of $j_c$ is symmetric about the $z$ axis
[$j_c(-\theta)=j_c(\theta)$], and the asymmetry of the $H_z$
profiles is caused by an inclined applied magnetic field. However,
such asymmetry can also result from the ``opposite'' situation
when the applied field is along the $z$ axis while pinning is not
symmetric about this axis. It is this situation that occurs when
columnar defects are introduced at an angle $\theta_1$ to the $z$
axis. In this case $j_c(\theta)$ in the interval $|\Delta
\theta|\equiv |\theta -\theta_1| < \theta_t$ is larger than
outside this interval. The enhancement of $j_c$ in this interval
is caused by the zigzag structure of vortices when a part of their
length is trapped by strong columnar defects, while outside the
interval pinning by the columnar defects is ineffective. Since
flux lines are curved in the critical state, this enhancement of
$j_c$ leads to an asymmetry of the current-density distribution
across the thickness of the strip, and thus to a nonzero $\Delta
H_z$. We carry out the calculation of the asymmetry of the $H_z$
profiles at the upper surface of the strip and of the shift of the
d-line for the following dependence $j_c(\theta)$:
     \begin{eqnarray}  \label{13}  
j_c(\theta) =j_{c0}\left [ 1 + p_1\exp[- q_1(\tan\theta -
 \tan\theta_1)^2 ]\right ] \,,
     \end{eqnarray}
which models an increased flux-line pinning by columnar defects at
angles $\theta$ near $\theta_1$, Fig.~\ref{fig5}. Here $p_1$ and
$q_1$ are some positive dimensionless parameters ($q_1 \sim
\theta_t^{-2}$), and $j_{c0}$ is the current density in a sample
without columnar defects ($j_{c0}$ describes, e.g., pinning by
point defects). Although formulas (\ref{5}) and (\ref{7}) are
still valid for thin strips with such $j_c(\theta)$, these
formulas fail near the d-line, and so we carry out calculations of
$\Delta H_z$ here, using the numerical solution \cite{EH1} of the
two-dimensional critical state problem for a strip of finite
thickness. In Fig.~5 we show the current and magnetic-field
distributions in the strip with pinning described by
Eq.~(\ref{13}), while in Fig.~\ref{fig6} the $H_z$ profiles and
the shift of the d-line are presented.

In Fig.~6 the decrease of the shift with increasing $H_a$ can be
qualitatively explained as follows: The characteristic angle
$\theta$ of the {\it curved} flux lines in the strip is of the
order of $j_{c0}d/H_z$ [i.e., for most of the flux-line elements
$\theta$ lies in the interval $-(j_{c0}d/H_z)< \theta <
(j_{c0}d/H_z)$]. When the external magnetic field increases, this
angle decreases and tends to zero. Thus, the flux-line pinning
(the current distribution) becomes practically uniform across the
thickness of the sample at sufficiently large $H_{az}$, and the
shift vanishes. These considerations are valid for any type of
anisotropic pinning, but there is one more reason for the
disappearance of the shift in samples with columnar defects: When
$\theta_1>\theta_t$ and $j_{c0}d/H_z$ is less than
$\theta_1-\theta_t$, the zigzag structure of vortices disappears
in the sample, and the flux-line pinning by these defects becomes
ineffective. The trapping angle determining the width of the peak
in $j_c$ decreases with increasing $H$
as some power of $H_{\Phi}/H$ if the
applied field is of the order of the matching field $H_{\Phi}$
($H_{\Phi}$ is a measure of the density of the columnar defects).
\cite{bl1} Thus, if $\theta_t>\theta_1$ at $H=0$, the
disappearance of the shift can occur at some magnetic field
associated with $H_{\Phi}$. This is just observed in the
experiment. \cite{t} Note that in this case the disappearance of
the shift is due to the disappearance of the zigzag structure of
vortices in the superconductor rather than to the loss of the
interlayer coherence.

Finally, we emphasize the unresolved problem of the analysis
presented in this section: The shift calculated for the
experimental ratio \cite{t} $(d/w)\sim 0.06$ is noticeably less
than the experimentally observed shift ($\Delta H_z$ decreases
with decreasing $d$). A variation of the parameters $p_1$ and
$q_1$ in Eq.~(\ref{13}) cannot change this conclusion. For
example, although the maximum value of the function $|x_{\rm
min}(H)|$ increases with $p_1$, it tends to a limit of the order
of $(d/2)\tan \theta_1$ at $p_1 \gg 1$; see Fig.~6.

\section{Conclusions}

We have considered $H_z$ profiles at the upper surface of a {\it
thin} strip whose thickness $d$ is much less than its width $2w$.
To the leading order in $d/w$, these profiles are symmetric in
$x$, $H_z(-x,d/2)=H_z(x,d/2)$. However, the analysis to the next
order in this small parameter reveals the asymmetry of the
profiles, $H_z(x,d/2)-H_z(-x,d/2)$, which coincides with $\Delta
H_z(x) \equiv H_z(x,d/2)-H_z(x,-d/2)$, the difference of $H_z$ at
the upper and lower surfaces of the strip. We calculate $\Delta
H_z(x)$ and show that this $\Delta H_z$ differs from zero in an
oblique magnetic field ${\bf H}_a=(H_{ax},0,H_{az})$ and depends
on the scenario of switching on this field. For definiteness, we
analyze in detail the so-called third scenario \cite{obl1} when
$H_{ax}$ is switched on before $H_{az}$.

In the region $|x|<a$ where the flux-free core occurs [where the
symmetric part of $H_z(x)$ is almost equal to zero], the asymmetry
of the $H_z$ profiles exists even for the ${\bf H}$-independent
$j_c$ (the Bean model) and is due to the asymmetric shape of the
flux-free core in the oblique magnetic field. Outside this region
($|x|>a$) the asymmetry appears only if $j_c$ depends on the
magnitude of the local magnetic induction or if there is an
out-of-plane anisotropy of $j_c$. The asymmetry of the magnetic
field profiles in oblique magnetic fields was observed in
Ref.~\onlinecite{w}; see also Fig.~3e in Ref.~\onlinecite{I}.

An asymmetry of the $H_z$ profiles also appears when the flux-line
pinning is not symmetric about the normal to the strip plane
(about the $z$ axis). This situation occurs when inclined columnar
defects are introduced into the strip. In this context we have
discussed the experimental data of Itaka et al. \cite{t} It is
shown that although these data can be qualitatively understood
from our results, there is a quantitative disagreement between the
experimental and theoretical results on the d-line shift.

\acknowledgments

This work was supported by the German Israeli Research Grant
Agreement (GIF) No G-705-50.14/01.

\appendix

\section{Formulas for $H_z$ at the surfaces of the strip} 

Using the Biot-Savart law, the $H_z$ component of the magnetic
field at the point ($x_0$,$z_0$) of the strip can be written in
the form:
\begin{eqnarray} \label{A1}
  H_z(x_0,z_0) = H_{az}\!+\!{1\over 2\pi}\!
 \int_{-d/2}^{d/2}\!\!\!\!\!\!dz\! \int_{-w}^{w}\!\!\!\!\!\!
 dx\,{j_y(x,z) (x\!-\!x_0)\over r^2} \,.\
\end{eqnarray}
where $r^2=(x-x_0)^2+(z-z_0)^2$. Using the smallness of the ratio
$d/w$, we now simplify this formula. Let us consider the integral
\begin{eqnarray} \label{A2}
 Q(z,x_0,z_0) =\!\! \int_{-w}^{w}\!\!\!\!\!\!
 dx\,j_y(x,z)\!\left[ {(x-x_0)\over r^2}-{1\over (x-x_0)} \right]
\end{eqnarray}
which appears if one calculates the difference between the
expression (\ref{A1}) and formula (\ref{2}) for the infinitely
thin strip. The main contribution to this $Q(z,x_0,z_0)$ is
determined by the $x$ values near $x_0$, $|x-x_0| \sim d$. Since
the current density $j_y(x,z)$ in the critical state of the strip
changes in the $x$ direction on a scale which considerably exceeds
$d$, in the calculation of $Q(z,x_0,z_0)$ we may use the
expansion, $j_y(x,z)\approx j_y(x_0,z)+(x-x_0)j_y'(x_0,z)$ where
$j_y'(x_0,z)\equiv \partial j_y(x_0,z)/\partial x_0$. Inserting
this expansion into integral (\ref{A2}), we find that
\begin{eqnarray}\label{A3}
 Q(z,x_0,z_0)\! \approx \! -\pi j_y'(x_0,z)|z\!-\!z_0|\! +\!
 j_y (x_0,z) {\rm O}({d^2\over w^2}), ~~
\end{eqnarray}
and the last term in this expression may be omitted. Putting $z_0
= \pm d/2$, we find
\begin{eqnarray} \label{A4}
 {1\over 2\pi}\!\int_{-d/2}^{d/2}\!\!\! Q(z,x_0,\pm {d\over 2})dz =
 - {dJ\over dx_0}{d\over 4}\pm {1\over 2} \Delta H_z(x_0) \,,
\end{eqnarray}
where $\Delta H_z(x)$ is given by formula (\ref{5}), and $J(x)$ is
the sheet current,
\[
 J(x)\equiv \int_{-d/2}^{d/2} j_y(x,z)dz .
\]
Eventually we arrive at
\begin{equation} \label{A5}
 H_z^{\pm}(x)\!=\!H_{az}+\!{1 \over 2\pi}\!\!\int_{-w}^w\!\!
 { J(t)\, dt \over t-x} - {d J(x) \over
 dx}{d\over 4} \pm {\Delta H_z(x)\over 2} \,.
\end{equation}
The first two terms in Eqs.~(\ref{A5}) yield $H_z(x)$ for the
infinitely thin strip, while the third and forth terms are
corrections to this result due to the finite thickness of the
strip. These corrections are relatively small (of the order of
$d/w$).

The above derivation of Eq.~(\ref{A4}) fails  in the region of the
strip, $|x|\le a$, where a flux-free core occurs in the sample.
The boundary of the core, $x_f(z)$ [or equivalently $z_f(x)$], can
be calculated from $J(x)$ at $|x|<a$; see Ref.~\onlinecite{1}. In
this region, the flux lines are practically parallel to the
surfaces of the sample, and they are sandwiched between the
surfaces and the core. If $|x_0-x_f(z_0)| \lesssim d$, i.e., if
the point ($x_0$, $z_0$) lies near the boundary of the flux-free
core, one cannot transform Eq.~(\ref{A2}) into Eq.~(\ref{A3}).
However, at $|x|<a$ we can repeat the above analysis, integrating
over $z$ rather than over $x$ in expression (\ref{A2}). If
$j_y(x,z)$ is almost independent of $z$ in the region between the
core and the surfaces of the sample, we eventually arrive at the
same formula (\ref{A4}) for $|x|< a$. This formula fails only in
the vicinity of the points $x=\pm a$ which are the positions of
the flux front for the infinitely thin strip and near the points
$x=0$, $x=\pm w$.

 \begin{figure}  
\includegraphics[scale=.44]{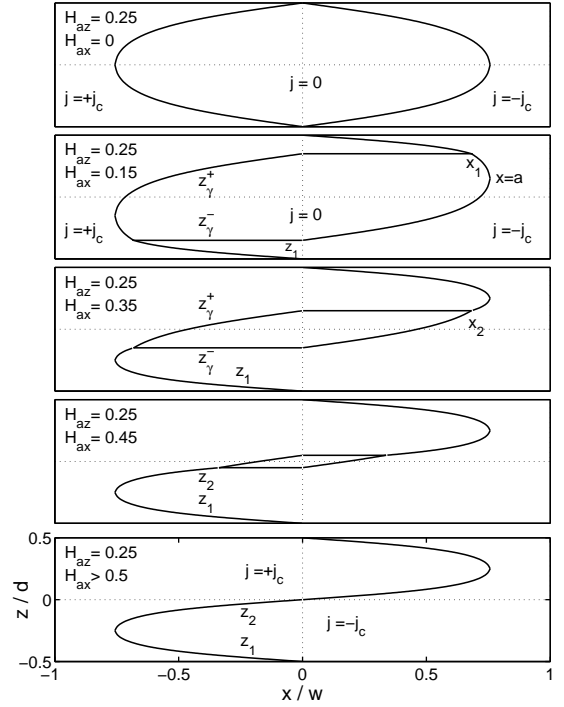}           
\caption{\label{figA} The current and flux fronts in a thin strip
to which first the in-plane magnetic field component $H_{ax}$ is
applied and then the perpendicular $H_{az}$ (scenario 3). Shown
are the fronts for $H_{ax}= 0$, 0.15, 0.35, 0.45 and 0.5 (or
larger), and $H_{az} = 0.25$ in units of $j_c d$. The fronts for
other values of $H_{az}$ are scaled with $a$. The lines
$z_{\gamma}^+(x)$, $z_{\gamma}^-(x)$, $z_1(x)$, $z_2(x)$ are
described in Appendix B.
 } \end{figure}  

\section{Flux-free core for the third scenario}  

In Ref.~\onlinecite{mbi} for a thin strip in an oblique magnetic
field, the shapes of the flux-free core and of the lines
separating regions with opposite directions of the critical
currents were presented only for two scenarios of switching on the
external magnetic field (1: at constant angle $\theta_0$, 2: first
$H_{az}$ then $H_{ax}$). Below we present the corresponding
formulas for the third scenario (when $H_{ax}$ is applied before
$H_{az}$). As in Ref.~\onlinecite{mbi}, we shall use the Bean
model here, i.e., we assume that $j_c$ is constant. However, the
formulas of this Appendix remain true also for the case
$j_c=j_c(\theta)$ where $\theta$ is the angle between the local
direction of ${\bf H}$ and the $z$ axis; see below.

Let $x=\pm a$ be the positions of the flux front in the infinitely
thin strip. Within the Bean model, the sheet current
$J(x)$ at $-a \le x \le a= w/{\rm cosh}(\pi H_{az}/j_cd)$ is
\cite{B1,B2}
\begin{equation} \label{B6}
 J(x)=-{2\over\pi}J_c\, {\rm arctan}
    {x \sqrt{w^2-a^2} \over \sqrt{a^2-x^2}} \,.
\end{equation}
In Fig.~\ref{figA} we show the lines $z_1(x)$ and $z_2(x)$
separating the regions with opposite directions of the critical
currents and the flux-free core composed of the lines
$z_{\gamma}^+(x)$ and $z_{\gamma}^-(x)$. Using the method of
Ref.~\onlinecite{mbi}, we find the formulas that describe
$z_1(x)$, $z_2(x)$, $z_{\gamma}^+(x)$ and $z_{\gamma}^-(x)$ at
$-a\le x \le 0$. The appropriate formulas for $a \ge x \ge 0$ can
be obtained by the substitution: $x\to -x$, $z\to -z$.

When $H_{ax}\le j_cd/4$ and $-x_1\le x \le 0$, one has
\begin{eqnarray}
z_{\gamma}^+(x)={d\over 2}-{H_{ax}\over j_c}- {J(x)\over 2j_c},
\label{b2} \\
z_{\gamma}^-(x)={H_{ax}\over j_c}- {d\over 2}, \label{b3} \\
z_1(x)={J(x)\over 4j_c}- {d\over 2}, \label{b4}
\end{eqnarray}
while if $-a\le x \le -x_1$, the $z_{\gamma}^+(x)$ is still given
by Eq.~(\ref{b2}), but
\begin{eqnarray}\label{b5}
z_{\gamma}^-(x)={J(x)\over 2j_c} -{H_{ax}\over j_c}- {d\over 2}.
\end{eqnarray}
Here $J(x)$ is the sheet current at the point $x$, and the point
$x_1$ follows from the condition $|J(x_1)|=4H_{ax}$.

When $j_cd/4 \le H_{ax}\le j_cd/2$ and $-x_2 \le x \le 0 $, the
functions $z_{\gamma}^+(x)$, $z_{\gamma}^-(x)$, $z_1(x)$ are
described by formulas (\ref{b2})-(\ref{b4}) where the point $x_2$
is determined by the condition $|J(x_2)|=2j_cd-4H_{ax}$. At $-a
\le x \le -x_2$ only the lines $z_1(x)$, $z_2(x)$ exist; the line
$z_1(x)$ is given by Eq.~(\ref{b4}), while
\begin{equation}\label{b6}
z_2(x)=-{J(x) \over 4j_c}.
\end{equation}

For high values of $H_{ax}$, when $j_cd/2 \le H_{ax}$, the
flux-free core disappears, and one has Eq.~(\ref{b4}) for $z_1(x)$
and Eq.~(\ref{b6}) for $z_2(x)$ at $-a\le x \le 0$.

Using the formulas of this Appendix and Eq.~(5), it is easy to
calculate $\Delta H_z(x)$ in the region $|x|\le a$. One has
\begin{eqnarray}
\Delta H_z(-a\le x \le x_1)=-{H_{ax}\over j_c}{dJ\over dx}, \\
\Delta H_z(-x_1\le x \le 0)=-{0.25J+H_{ax}\over 2j_c}{dJ\over dx},
\end{eqnarray}
at $H_{ax}\le j_cd/4$,
\begin{eqnarray}
\Delta H_z(-a\le x \le x_2)=-{d\over 4}\left ({dJ\over dx}\right ), \\
\Delta H_z(-x_2\le x \le 0)=-{0.25J+H_{ax}\over 2j_c}{dJ\over dx},
\end{eqnarray}
at $j_cd/4 \le H_{ax} \le j_cd/2$, and Eq.~(B9) at $j_cd/2 \le
H_{ax}$ in the whole interval $-a \le x \le 0$. When $0\le x \le
a$, one can use $\Delta H_z(x)=-\Delta H_z(-x)$.

The formulas of this Appendix are applicable to the case
$j_c=j_c(\theta)$ since at $|x|\le a$, the flux lines are
practically parallel to the strip plane and $\theta \approx
\pi/2$. Hence, it is sufficient to put $j_c=j_c(\pi/2)$ in the
above formulas and to use $J(x)$ obtained numerically from the
appropriate solution of the critical state problem for the
infinitely thin anisotropic strip.

\section{$j_c(\theta)$ and $J_c(H_z,H_{ax})$ in terms of
$J_c(H_z,0)$}  

When $j_c$ depends only on $\theta$, one can reconstruct this
dependence from the $J_c(H_z)$ obtained at $H_{ax}=0$:
\cite{1,obl1}
     \begin{eqnarray} \label{2a}
 j_c(\theta) d = J_c(H_z) -H_z{d J_c(H_z) \over d H_z} \,, \nonumber \\
  \tan \theta   = { J_c(H_z) \over 2 H_z } \,.
     \end{eqnarray}

Inserting this parametric form of $j_c(\theta)$ into
Eq.~(\ref{1a}), we arrive at equations determining $J_c(H_z)$ at
$H_{ax}\neq 0$, \cite{obl1}
     \begin{eqnarray}  \label{10}
 {0.5J_c(H_z,H_{ax})+H_{ax}\over H_z}&=&{J_c(t_+,0)\over 2t_+},
 \nonumber \\
 {|0.5J_c(H_z,H_{ax})-H_{ax}|\over H_z}&=&{J_c(t_-,0)\over 2t_-},
 \nonumber \\
 {1\over H_z}&=&{1\over 2t_+}+\sigma {1\over 2t_-},
     \end{eqnarray}
where $\sigma$ is the sign of $[0.5J_c(H_z,H_{ax})-H_{ax}]$,
$J_c(H_z,H_{ax})$ denotes $J_c(H_z)$ at a given value of $H_{ax}$,
and hence $J_c(t,0)$ is the sheet current at $H_{ax}=0$. At
$H_{ax}=0$ one has $t_+=t_-=H_z$, and Eqs.~(\ref{10}) reduce to
$J_c=J_c(H_z,0)$ as it should be. From the three Eqs.~(\ref{10})
for the three unknown variables $J_c$, $t_+$, $t_-$ ($t_+$ and
$t_-$ are auxiliary variables) one finds $J_c(H_z)$ at $H_{ax}\neq
0$. The magnetic field profiles in oblique applied field are then
obtained by inserting this effective law $J_c(H_z)$ into the
critical state equations for the infinitely thin strip,
Eqs.~(\ref{2}) - (\ref{C4}).

{}


\begin{thebibliography}{}
\bibitem{c1} The idea of the imbalance of the currents
(see Fig.~4c in Ref.~\onlinecite{t}) leads to ``opposite'' current
distributions for the upper and lower surfaces of the sample.
Since the currents flowing on both these surfaces almost equally
generate $H_z$, the asymmetry of the $H_z$ profiles was
considerably overestimated in Ref.~\onlinecite{t}.
\end{thebibliography}
\end{document}